\def\boxit#1{\vbox{\hrule\hbox{\vrule\kern3pt
\vbox{\kern3pt#1\kern3pt}\kern3pt\vrule}\hrule}}

\def\expect#1{{ \langle{#1}\rangle }}

\def\frac#1#2{{#1}\,/{#2}}
\def\psid{{\psi^\dagger}}
\def\chid{{\chi^\dagger}}
\def\Bv{{\bf B}}
\def\Ev{{\bf E}}

\def\quarter{{\frac{1}{4}}}

\def\Dv{{\bf D}}
\def\DIV{{\bf \nabla}}
\def\g5{{\gamma_5}}

\def\beginEq{\begin{equation}}
\def\endEq{\end{equation}}
\def\beginEqarray{\begin{eqnarray}}
\def\endEqarray{\end{eqnarray}}
\def\labelEq#1{\label{eq:#1}}

\def\nl{\nonumber \\}

\def\NNRQED{{\rm NNRQED}}

\def\Lag{{\cal L}}

\def\G2b{\overline{G^2}}

\def\G{{\cal G}}

\def\pn1p{{(n+1)}}

\def\Dv{{\bf D}}
\def\sigmav{\mbox{\boldmath$\sigma$}}
\def\Ev{{\bf E}}
\def\Bv{{\bf B}}
\def\psid{{\psi^\dagger}}

\def\quarter{\mbox{$\frac{1}{4}$}}

\documentstyle{article}
\begin{document}
\title{ NRQED  IN BOUND STATES: \\
 APPLYING RENORMALIZATION TO AN EFFECTIVE FIELD THEORY}
\author{PATRICK LABELLE\thanks{E-mail address : pat@beauty.tn.cornell.edu.}\\
\small Newman Laboratory of Nuclear Studies, Cornell University, Ithaca,\\
NY  14853, USA}
\vspace{0.3cm}
\maketitle
\setlength{\baselineskip}{2.6ex}
\begin{center}
\parbox{13.0cm}
{\begin{center} ABSTRACT \end{center}
{\small \hspace*{0.3cm} This is a pedagogical introduction
 to NRQED, a nonrelativistic approximation to
quantum electrodynamics. It can be made to reproduce
 QED to any degree of precision when applied to
nonrelativistic systems and is particularly powerful when applied to bound
states. We will begin by explaining what makes bound states so special, and
so difficult to treat using QED. We will then develop the most  na\"\i ve
nonrelativistic approximation of QED which will be shown to fail when
applied to a simple example (hyperfine splitting in positronium). We will
see what was missing to make the theory well-defined, and the resulting
theory will be, essentially, NRQED. We will conclude by discussing what
makes NRQED so much more powerful than QED in nonrelativistic bound states,
and what the approach we followed can teach us about ``new physics" beyond
QED.}}
\end{center}

\section{Introduction}

\par Quantum electrodynamics is without any doubt the most successful theory
developed so far to deal with the behavior of elementary particles. It
encompasses in an elegant and covariant framework the interactions of
charged particles with electromagnetic fields. It is unfortunately
overly sophisticated when it comes to dealing with nonrelativistic
situations, since it does not take advantage of the simplifications
possible under these circumstances. This is true for any nonrelativistic
calculation, but is even more striking in the case of bound states
calculations.

Indeed, although most issues of principle concerning nonrelativistic bound
states have long been resolved, there remain significant technical problems
connected with the study of these systems. Central among these is the
problem of too many energy scales. Typically, a nonrelativistic system
has at least three important energy scales: the masses $m$ of the
constituents, their three-momenta $p \sim mv$, and their kinetic energies
$K \sim mv^2$. These scales are widely different in a nonrelativistic
system since $v \ll 1$ (where the speed of light $c=1$), and this
complicates any analysis of such a system.

To illustrate the problems that arise, consider a traditional
Bethe-Salpeter\cite{BS}  calculation of the energy
 levels of positronium. The potential
in the Bethe-Salpeter equation is given by a sum of two-particle irreducible
Feynman amplitudes. One generally solves the problem for some approximate
potential and then incorporates corrections using time-independent
perturbation theory. Unfortunately, perturbation theory for a bound state
is far more complicated than perturbation theory for, say, the electron
$g$-factor. In the latter case a diagram with three photons contributes
only in order $\alpha^3$. In positronium a kernel involving the exchange
of three photons  can also contribute to order
$\alpha^3$, but the same kernel will contribute to all higher orders
as well:
\smallskip
\begin{equation}
 \expect{V_1}~=~\alpha^3 m \bigl( a_0 + a_1 \alpha + a_2 \alpha^2+ ...
\bigr)
\end{equation}
\smallskip
So in the bound state calculation there is no simple correlation
between the importance of an amplitude an the number of photons in it.
Such behavior is at the root of the complexities in high-precision
analyses of positronium or other QED bound states, and it is a direct
consequence of the multiple scales in the problem. Any expectation
value like Eq.(1) will be some complicated function of the ratios
of the three scales in the atom:
\smallskip
\begin{equation}
 \expect{V_1}~=~\alpha^3 m~ F(\expect{p}/m,\expect{K}/m).
\end{equation}
\smallskip
Since $\expect{p}/m \sim \alpha$ and $\expect{K}/m \sim \alpha^2$, a Taylor
expansion of $F$ in powers of these ratios generates an infinite series
of contributions just as in Eq.(1).  A similar series does not
occur in the $g$-factor calculation because there is but one scale
in that problem, the mass of the electron.

Traditional methods for analyzing these bound states fail to take
advantage of the nonrelativistic character of these systems; and
atoms like positronium are very nonrelativistic: the probability for
finding relative momenta of ${\cal O}(m)$ or larger is roughly $
\alpha^5 \simeq 10^{-11}$!

There exists however  an alternative formulation of QED that corresponds
closely to to the usual nonrelativistic theory, but where relativistic
and radiative corrections can be systematically included. This new
formulation is called nonrelativistic QED or NRQED \cite{Caswell}
and greatly simplifies calculations involving nonrelativistic
systems, especially bound states.

Let us also point out that the difficulties associated with the
many different energy scales in bound states are not limited to QED. Consider
the study of nonrelativistic QCD bound states, like the $\Upsilon$
for example. Even if the means of extracting the properties of such
systems, namely simulating the system on a lattice, are very different than the
use of the Bethe-Salpeter equation, the nonrelativistic nature of
the bound state still greatly complicates  calculations. Indeed, the
space-time grid used in such a simulation must accommodate wavelenghts
covering all of the scales in the meson, ranging from $1/mv^2$ down
to $1/m$. Given that $v^2 \sim 0.1$ in $\Upsilon$, one might easily
need a lattice as large as 100 sites on side to do a good job. Such a lattice
could be three times larger than the largest wavelenght, with a grid spacing
three times smaller than the smallest wavelenght, thereby limiting the errors
caused by the grid. This is a fantastically large lattice by contemporary
standards and is quite impractical.
 The strong force equivalent of NRQED,$\,$  NRQCD\cite{NRQCD}, greatly reduces
the size of the lattices required to study heavy quarks systems
 since it essentially eliminates the mass of the
meson as a relevant energy scale in the calculation (we will soon see
how this occurs in NRQED).

In the next section, we will develop a na\"\i ve  nonrelativistic approximation
of QED, which we will refer to as ``NNRQED" to emphasize the fact that this
theory is {\it not} NRQED as described in Ref[1].
 It is in fact not well defined, as we will show
later. Only when we will make it well-defined with the help of the concept
of renormalization will it become  NRQED.

\section{NNRQED as a low energy theory}

To take advantage of the simplifications associated with nonrelativistic
systems, one must approximate QED in the limit $p/m \ll 1$ (which is
equivalent to the limit $v/c \ll 1$). One way
to do this consists of simply expanding the QED Lagrangian  in powers
of $p/m$ to obtain NNRQED. However, it is possible to define
NNQRED without prior knowledge of QED.
Indeed, one can build up the Lagrangian
of NNRQED by imposing restrictions coming from the symmetry obeyed by the
theory, such as gauge invariance,  chiral symmetry in the limit $m_e
\rightarrow
0$, and Lorentz invariance for the photon kinetic term. Lorentz invariance
for the rest of the Lagrangian and renormalizability are {\it not} necessary.

It is of course possible to write down an infinite number of terms respecting
these symmetries,
but this is not a problem since operators of dimension greater than $4$ will
be multiplied by coefficients containing inverse powers of $m$, the only
available energy scale in the problem. But since we are considering the
limit  $p/m \ll 1$, these operators  will yield
contributions suppressed by that many powers of $p/m$ and therefore,
for a given accuracy, only a few terms need to be kept in an actual
calculation. The first few terms one obtains are then
\beginEqarray
\Lag_{\NNRQED} & = &  - \quarter(F^{\mu\nu})^2 +
  \psid \biggl\{ i\partial_t - e A_0 + \frac{\Dv^2}{2m} + \frac{\Dv^4}{8m^3}
   \nl
&& + c_1~\frac{e}{2m}\,\sigmav\cdot\Bv + c_2\frac{e}{8m^2}\,\DIV \cdot \Ev \nl
&& + c_3~\frac{ie}{8m^2} \,\sigmav\cdot(\Dv\times\Ev - \Ev\times\Dv)
   + ...\biggr\} \psi \nl
&& + \frac{d_1}{m^2}\,(\psid\psi)^2 + \frac{d_2}{m^2}\,(\psid\sigmav\psi)^2
 +... \nl
 && + \mbox{~positron and positron-electron terms}.
 \labelEq{nrlag}
 \endEqarray
where $\Dv$ is the gauge-invariant derivative. An example of a
 positron-electron term  (and the one we will focus on  later)
 is $ d_3/m^2 \,
(\psid\sigmav \chi) \cdot (\chid\sigmav \psi) $, which represents, to
lowest order, the process $e^+ e^- \rightarrow e^+ e^- $ in the $s$
channel. This term, and the other two having coefficients starting with
$d$'s are examples of ``contact" interactions.

 Notice  that
we have  recovered the interactions familiar from nonrelativistic quantum
mechanics such as the first order relativistic correction to the
kinetic energy ($ \Dv^4/8m^3$),$~$ the spin-orbit interaction$~~~$
($ c_1~e/2m\,\sigmav\cdot\Bv$),$~$
 and the Darwin term ($ c_2~e/8m^2\,\DIV \cdot \Ev$).

We now have to fix the coefficients appearing in $\Lag_{\NNRQED}$\footnote{
Had we chosen to expand the QED Lagrangian, these coefficients would already
be fixed. }. To do so, we may simply compare tree scattering amplitudes in both
QED and NNRQED, at a given kinematic point (which we choose to be at threshold,
{\it i.e.} with the external particles at rest). This fixes $d_1$ in  Eq.(3)
  to be equal to $-e^2/4m^2$ for example.

\section{Bound states diagrams in NNRQED}

\subsection{Introduction}

Now that NNRQED is completely defined, we can apply it to bound state
calculations, where its usefulness is  most apparent. To be specific,
consider the energy levels of positronium. As noted earlier, the expansion
in Feynman diagrams of NNRQED reduces to  conventional Rayleigh-Schr\"odinger
perturbation theory. In a bound state, the ``external" wavefunctions
of course don't correspond to free wavepackets but to bound states.
In the traditional Bethe-Salpeter formalism, the form of these wavefunctions
depend on which interactions one desires to include exactly, and which
interactions
one wants to treat perturbatively.
 Depending on this choice, the external wavefunctions
may be as simple as the Schr\"odinger wavefunctions or as complicated as the
Dirac wavefunctions.  The most useful gauge for nonrelativistic systems
is the Coulomb gauge. That gauge  permits us to keep
 the simplest interaction for the
zeroth order kernel, namely the Coulomb interaction, in which case
the Bethe-Salpeter equation reduces to the Schr\"odinger equation
and the wavefunctions are the ones found in any textbook on quantum mechanics.
In this calculation we will work only with $1S$ wavefunction in positronium
(reduced mass $\mu=m/2$) which is given, in momentum space, by
\smallskip
\begin{equation}
 \Psi(\vec p)~=~ { 8 \pi^{1/2} \gamma^{5/2} \over
(\vec p^2 + \gamma^2)^2}
\end{equation}
\smallskip
where $\gamma$ is the typical bound state momentum equal to $\mu v
 = m \alpha /2$ in positronium. The ground state energy is $- \gamma^2/2 \mu =
- m \alpha^2/4 $. Physically, an infinite number of Coulomb interactions is
incorporated in the wavefunction.

 The reason why at least an infinite number  of
Coulomb kernels must be taken into account in the wavefunction can be
easily understood by considering  Fig.[1a], where a generic bound state
diagram is shown. $K_1$ and $K_2$ represent some unspecified kernels.
Let us now add one  Coulomb interaction between the two kernels to yield
Fig.[1b]. In conventional
scattering theory, one would conclude that this diagram is of higher
order than  diagram [1a]  because of
the factor $\alpha$ coming from the vertices. However, this argument does
not hold in a bound state, because the typical momentum flowing
through the Coulomb interaction is of order $\gamma$. Indeed, consider
how the diagram changes when one adds the Coulomb interaction. The
integral acquires a new term of the form
\smallskip
\begin{equation}
 \int {d^3k \over (2 \pi)^3}~ {- e^2 \over \vert \vec k - \vec l \vert^2}~
{ 1 \over \ (- \gamma^2 - \vec k^2)/m } \simeq  \int d^3k { m \alpha
\over (\gamma^2 + \vec k^2) ( \vert \vec k - \vec l \vert^2)} ,
\end{equation}
\smallskip
where $1/ \vert \vec k - \vec l \vert^2$ is the Coulomb propagator
and $m/ (- \gamma^2 - \vec k^2)$ is the nonrelativistic propagator for
the $e^-e^+$ pair.

We see that the mass factors out (more about this later!), leaving $\gamma$
as the only energy scale in the integral. By dimensional analysis, the
final result will be of order
 $\simeq m \alpha / \gamma \simeq 1$. This shows that in a bound
state, adding a Coulomb interaction in fact does not  lead to a diagram
of higher order, which is why one must include all of them from the
start in the wavefunctions. This implies the relation expressed in Fig.[2]
which is in fact a diagrammatic representation of the Schr\"odinger
equation.

\subsection{Counting rules}

The argument we used to show that adding a Coulomb interaction does not
increase the order of a bound state diagram is  an example of a NNRQED
counting rule. These rules are extremely important since
 they permit one to estimate the order of
contribution of a diagram without  calculating it.
Let us first consider non-recoil diagrams\footnote{In the case of
a bound state having  equal mass
constituents, as  positronium, there is no distinction possible between
 ``recoil effects" and the so-called ``retardation effects". This is not, for
example, the
case in muonium or hydrogen where (sometimes
confusing) distinctions are made between
non-recoil and non-retardation approximations. What we mean here by the
 non-recoil approximation is the limit in which
 the transverse photon does not carry
any energy.}.
In that approximation, the counting rules are based on the observation that
the mass $m$ of the electron  always factors out of the integrals so that it
 no longer represents a relevant energy scale.
This is clear since $m$ factors out of the nonrelativistic fermion propagators,
as we saw in Eq.(5), and it enters the rules of the NNRQED vertices only
as an overall factor.
We therefore see that  the use
of NNRQED eliminates the relativistic energy scale of the system, which, in
the light of the discussion in the introduction,
 greatly simplifies the analysis.

Starting from this observation, we can recast the discussion of the Coulomb
interaction in the following manner: adding a Coulomb interaction between two
kernels as in Fig.[1] leads to an additional factor of $\alpha$ coming from
the vertices and a factor of $m$ coming from the additional fermion
propagator. The corresponding overall factor of $m \alpha$ must be cancelled
by a factor having the dimensions of energy in order to keep the dimensions
straight, but since the only energy scale left in the problem is $\gamma$,
we finally obtain a correction of order $m \alpha / \gamma \simeq 1 $. It
now becomes obvious that the Coulomb interaction is the only kernel
having the property of not increasing the order of a bound state diagram
as all the other interactions contain additional factors of $1/m$.
 Each of these $1/m$ factors will have to be cancelled by a corresponding
factor of $\gamma$, leading therefore to a result of higher order in
$ \alpha$.

With these rules, it becomes extremely simple to evaluate the order of the
contribution of a given diagram. All we need to know  is that the external
wavefunctions contribute  a factor $m \alpha^2$ to the energy (this
can be seen by inserting a simple Coulomb interaction between two wavefunction
and noticing that the result is of order $\alpha \gamma $). To obtain
the contribution of a given kernel, one has simply to count how many powers
of $\alpha$ and of $m$ this kernel has relative to the Coulomb interaction.
After cancelling the $1/m$'s  with factors of $\gamma$, the overall factor
of $\alpha$ gives the order of the diagram. For example consider inserting
a transverse photon (in the non-recoil limit)  between two wavefunctions
as  in Fig.[3]. The transverse photon vertex has the same power of
 $e$ as the Coulomb vertex, but it has an additional factor
 of $1/m$. Taking the two vertices into account, we can conclude that this
diagram will contribute to order $m \alpha^2 \times \gamma^2/m^2 \simeq
m \alpha^4$. The same is true of the annihilation diagram and of the
spin-spin interaction as they also contain a factor of
$1 /m^2$ relative to the Coulomb interaction. Another important observation
is that these diagrams will contribute to only one order in
 $\alpha$\footnote{We are still limiting ourselves to the non-recoil limit.},
  in contradistinction with Feynman diagrams in conventional Bethe-Salpeter
analysis.

The situation is only slightly more complicated when one takes into account
recoil effects.  We will not dwell on this issue here, but let us just
mention that they make the kinetic energies $K$ enter as a mass scale
and that they are at the origin of the appearance of log's of $\alpha$
which are characteristic of bound states.

\section{A calculation that goes wrong}

Now let's do a calculation! We will consider
 the hyperfine splitting\footnote{Which we will denote by
``hfs" in the following.} (E (triplet
state) - E (singlet state)) of ground state positronium. We first have to find
which diagrams will contribute to the lowest order hfs. Using the counting
rules described earlier, we easily find that there are only
two such diagrams, and that they will contribute to the order $m \alpha^4$.
One of these diagrams is, naturally,  the spin-spin interaction familiar from
quantum mechanics; the second interaction is the annihilation
 contact interaction which is of course absent in hydrogen.
For the sake of simplicity, we will focus our attention on
 the annihilation diagram. It will simplify the
 discussion without affecting the conceptual issues
we want to address.

As explained earlier, the annihilation interaction is represented by a contact
term; it has no momentum dependence and has for Feynman rule\footnote{This
is the Feynman rule for the {\it amplitude} \cal M, and not $-i$ \cal M, as
the usual QED Feynman rules correspond to. Also, to be consistent with the
normalization of the wavefunctions, the external spinors have a nonrelativistic
normalization, {\it i.e.} $\psi^\dagger \psi = 1$ for example.}
$e^2/4 m^2 \chi^\dagger \vec \sigma \psi \cdot \psi^\dagger \vec \sigma \chi
$. The spin average of this expression gives $2$ in the triplet
state and $0$ in the singlet state (as it must since the singlet state
cannot decay to one photon, by charge conjugation invariance).
Therefore, the contribution to the hfs coming from this interaction is
given by
\smallskip
\begin{equation}
\int {d^3 p \over (2 \pi)^3} { 8 \pi^{1/2} \gamma^{5/2} \over (\vec p^2 +
\gamma^2)^2}~2 \times {e^2 \over 4 m^2}~ \int {d^3 q \over
 (2 \pi)^3} { 8 \pi^{1/2} \gamma^{5/2} \over (\vec q^2 +
\gamma^2)^2} ~=~{ m \alpha^4 \over 4}
\end{equation}
\smallskip
which is indeed the correct result\cite{iz}.

The first correction beyond the tree approximation
 is of order $m \alpha^5$\cite{iz}.  To isolate it, we may apply
the counting rules explained earlier. They tell us that we need to
join to the annihilation interaction a kernel that will supply one
more power of $1/m$ (indeed, this will force the integral
to contribute an additional power of $\gamma$, resulting in a contribution
to the energy of order $m \alpha^5$). It turns out that there is {\it no} such
kernel. The smallest power of $1/m$ one can obtain is $1/m^2$, which
seems to indicate that our theory predicts the next order correction
to be of order $m \alpha^6$. Actually, the situation is even worse. To see
this, consider the simplest of these next order diagrams, the one corresponding
to two annihilation diagrams sewn together. The corresponding integral is
(I write  the external wavefunction integrals directly as $\vert \Psi(0)
\vert^2
$, as they decouple from the rest of the diagram):

\begin{eqnarray}
& \Biggl( \vert   \Psi(0) \vert^2 & Tr(\sigma_i \sigma_j)~ ({e^2 \over 4m^2})^2
\chi^\dagger \sigma_i \psi \psi^\dagger \sigma_j \chi ~ \int {d^3p \over (2
\pi)^3}~{1 \over -\gamma^2/m - \vec p^2/m} \Biggr) \Biggl|^{spin~1}_{spin~0}
\nonumber \\
  =~- & \vert   \Psi(0) \vert^2 & 4 ({e^2 \over 4 m^2})^2
 \int{d^3 p \over (2 \pi)^3} {m \over \vec p^2 + \gamma^2}
\end{eqnarray}

\noindent where we have used $~ Tr(\sigma_i \sigma_j)~=~2 \delta_{ij}$.

We immediately see that this integral is ill-defined, as it has a linear
UV divergence. One can of course argue that, since our theory is based
on  a nonrelativistic expansion, one must restrict the momentum integration
to $p \ll m$,
 but what cutoff should one
use? $m/2$, $m/10~$, $m/20~$? We realize that, as soon as we push the precision
beyond the tree approximation, our nonrelativistic theory is not well defined,
and it seems that we must bid farewell to all the nice simplifications inherent
to our nonrelativistic field theory...

\section{Renormalization  to the rescue }

Fortunately, all is not lost as  there is a
 well-known method that dictates us how we can make our
theory finite in a well-defined and entirely self-consistent manner. This
method is nothing other than renormalization.

To understand how renormalization can teach us how to make the nonrelativistic
theory well-defined, let us go back to conventional QED. In QED, to avoid
infinities, one regulates the theory by putting a cutoff $\Lambda$ on the
loops momentum. This throws away some physics but renormalization theory
tells us that this physics can be put back in the theory by shifting
(``renormalizing") the values of the QED parameters  and by adding
new, ``nonrenormalizable", interactions. There is an infinite string
of new interactions  which, by dimensional analysis, are accompanied
by increasing powers of $1/ \Lambda$. To fix the parameters, one calculates
a given process, at a given kinematic point, and compares with the
experimental value. For example, to fix $\alpha$ in QED, one might calculate
the form factor of the scattering of an electron by an external electromagnetic
field, in the long wavelenght limit. Once the parameters have been fixed,
the theory can be used to calculate any process, at any energy (as long as
it is small compared to the cutoff $\Lambda$ so that only a finite
number of the nonrenormalizable interactions must be taken into account
for a given accuracy). The conventional procedure is to take the limit
$\Lambda \rightarrow \infty$ at the end of the calculation so that
the nonrenormalizable interactions actually vanish. The price paid is,
as we know, that the original (``bare") parameters of QED become ill-defined
in that limit. We will discuss this point in more details in the conclusion.

Standard renormalization is readily adapted to NNRQED.
 The only difference here is that
there is a physical scale, $\approx m_e$, at
 which NNRQED is no longer applicable
and QED takes over. In this context, renormalization is telling us that
the physics taking place beyond $p \approx m_e$ can be put back in
NNRQED by an appropriate definition of its parameters, and by adding
new interactions representing an expansion in powers of $p/m$. In fact,
since our nonrelativistic Lagrangian contains already such an expansion,
the net effect of adding the cutoff is to change the (infinite) set
of NNRQED parameters. As long as we are considering processes in which the
external momenta are smaller than the electron mass, we are certain
to reproduce the QED result, up to a certain precision, with a
 finite number of diagrams providing we have defined the NNRQED
 parameters correctly.   To see how this works, let us go back to
 the annihilation
diagram. We must now think of  its coefficient as
 an expansion in $\alpha$. To fix
it to the order we are interested in, we must consider one loop
corrections to both NNRQED and QED, and require that  the scattering amplitudes
in both theories  agree (see Fig.[4]).  This must be done at a given
kinematic point, which, as already mentioned,  we choose to be
 at threshold (external
particles being at rest). As in QED, once the coefficients are fixed at
a given kinematic point, the theory can be used to calculate
 any process. The sum of the QED one-loop diagrams is
calculated to be $~-e^4 (1/2+1/9)/m^2 \pi^2~$\footnote{For the sake of
conciseness, we are side-stepping completely the question of
infrared divergences. Since we are working at threshold, there cannot
be any bremsstrahlung divergences, but there is a so-called ``Coulomb
singularity" divergence present. However, the Coulomb diagram of NNRQED
takes care completely of this problem, and this does not change anything
to our discussion.}. To match this result,
  the contact interaction must
acquire an additional finite piece to become $\chi^\dagger
\vec \sigma \psi \cdot \psi^\dagger \vec \sigma \chi \times
e^2/ 4 m^2 \{ 1  - e^2 (1+2/9)/\pi^2 \}$.  Then the lowest order
 calculation (Eq.(6))
becomes
\begin{equation}
\vert \Psi(0) \vert^2 \times \biggl\{ {e^2 \over 2 m^2}  -{e^4 \over m^2 \pi^2}
({1 \over 2} + {1 \over 9}) \biggr\} ~=~{ m \alpha^4 \over
 4} - {m \alpha^5 \over
\pi} (1 + {2 \over 9})
\end{equation}

The $\alpha^5$ correction is exactly what we needed
 to obtain the correct answer \cite{iz}!
Moreover, this procedure will take care of the divergence appearing in the
$\alpha^6$ terms as well. This can be seen by looking back at Fig.[4].
We see that the coefficient in front of the contact interaction will contain,
in addition to the constant term that we have just described and which comes
from the QED one-loop diagrams,  a series of NRQED one-loop diagrams
evaluated at threshold, which are infinite. It will, for example, contain
the scattering diagram representing the double annihilation
diagram of Eq.[7]. Therefore, to be rigorous, this term should
now be included in Eq.[8] and can be seen to play the role of a counterterm
which will render Eq.[7] finite\footnote{The precise way this occurs is that
the integrand $1/(\vec p^2 + \gamma^2)$ of Eq.[7] gets replaced by
 $1/(\vec p^2 + \gamma^2) - 1/\vec p^2$ which leads to a finite integral.
The counterterm integrand does not contain $\gamma$ as it comes from a
scattering amplitude.}.

\section{Conclusion}

With the coefficients of our nonrelativistic theory renormalized in the
way described above,  we have finally obtained a truly well-defined
nonrelativistic field theory, which is the NRQED described in refs[1].

We would like to conclude this paper by emphasizing two important aspects
of our discussion on NRQED. One of these aspects is conceptual in nature,
and has relevance far beyond the realm of nonrelativistic systems. The
 other is practical in nature and is at the heart of the power of NRQED
in bound states.

$\bullet$ It is probable that QED is ``wrong" in the sense that it must be
an approximate ``low-energy'' realization of a more fundamental theory.
The way in which this clearly shows up is the appearance of the well-known
UV divergences. Physically, these divergences tell us that we are pushing
the integration momenta far beyond the region of applicability of QED, in
a region where the ``complete" theory would have to be used. In that
respect, we see that the situation of QED with respect to the complete
theory is analogous to the situation of NRQED with respect to QED. From
that point of view, one should not, in QED, take the limit
$\Lambda \rightarrow \infty$. As in NRQED, there is a {\it physical}
finite value of the cutoff $\Lambda_{NP}$ signaling the presence of
new physics. This implies that the nonrenormalizable interactions
cannot be dispensed of. Therefore, at energies of order $\Lambda_{NP}$,
one would observe a complete breakdown of QED as the infinite
string of new interactions would contribute to ${\cal O}(1)$. However,
 experiments at energies as high as $\Lambda_{NP}$
are not necessary to probe the new physics. Low energy, but very accurate,
experiments can also teach us something about $\Lambda_{NP}$.
For example, there should be an interaction
of the form $ e \, m_e \bar \Psi F^{\mu\nu} \sigma^{\mu \nu}
 \Psi / \Lambda_{NP}^2$
which would contribute to the electron anomalous magnetic moment by
an amount proportional to $m_e^2/\Lambda_{NP}^2$.
The fact that theory agrees with experiment to 12 decimal places indicates
that $\Lambda_{NP} > 10^6 m_e \approx 1 TeV$.

$\bullet$ NRQED is an extremely powerful device when applied to
nonrelativistic bound states. To understand why, notice that the contributions
from relativistic states ({\it i.e.} QED diagrams) enter the theory only
by renormalizing the coefficients of the nonrelativistic interactions, and
this occurs when one matches the  {\it scattering} amplitudes of  NRQED to
the ones of QED. At that stage, no bound state analysis enter the problem.
It is only when these coefficients are defined that one attacks the
bound state part of the calculation, which, therefore, involves only
nonrelativistic interactions. We see that NRQED has {\it decoupled}
the high momentum contribution from the low-momentum one. This is the
fundamental ingredient in the usefulness of NRQED.

\bigskip
\section{Acknowledgments}
I would like to thank the organizers of the XIV MRST meeting for a very
stimulating and enjoyable experience.

I am indebted to my adviser, G.Peter Lepage, for his constant support
and $\infty$ patience in teaching me NRQED.
I also thank Lisa Angelos for many useful comments concerning this manuscript.

This work was supported by the Fonds  FCAR (Qu\'ebec)
and by a ``67" fellowship from NSERC.

\bigskip

\bibliographystyle{unsrt}

\end{document}